\documentclass[final,5p,twocolumn]{elsarticle}
\usepackage{graphicx}
\usepackage{epsfig}
\usepackage{amssymb}
\usepackage{amsthm}
\usepackage{float}
\usepackage[font={footnotesize,it}]{caption}

\journal{Physics Letters B}

\begin{document}

\begin{frontmatter}

\title{Thermodynamic geometry of holographic superconductors}

\author[label1]{Sayan  Basak}
\address[label1]{Department of Physics and Astronomy, Purdue University, 525 Northwestern Avenue,\\ West Lafayette, Indiana 47907-2036,United States}
\ead{basak@purdue.edu}

\author[label2]{Pankaj Chaturvedi\corref{cor1}}
\ead{cpankaj@iitk.ac.in}
\author[label2]{Poulami Nandi }
\ead{ polulamin@iitk.ac.in}
\author[label2]{Gautam Sengupta}
\ead{ sengupta@iitk.ac.in}
\address[label2]{Department of Physics, Indian Institute of Technology Kanpur, Kanpur 208016, India}
\cortext[cor1]{corresponding author}

\begin{abstract}

We obtain the thermodynamic geometry of a (2+1) dimensional strongly coupled quantum field theory at a finite temperature in a holographic set up, through the gauge/gravity correspondence. The bulk dual gravitational theory is described by a (3+1) dimensional charged AdS black hole in the presence of a massive charged scalar field.  The holographic free energy of the (2+1) dimensional strongly coupled boundary field theory is computed analytically through the bulk boundary correspondence. The thermodynamic metric and the corresponding scalar curvature is then obtained from the holographic free energy. The  thermodynamic scalar curvature characterizes the superconducting phase transition of the boundary field theory.

\end{abstract}


\end{frontmatter}

\section{Introduction}

The gauge theory/gravity correspondence has been one of the most significant advances in the study of the physics of fundamental forces. This holographically relates a weakly coupled (d+1) dimensional bulk classical theory of gravity coupled to matter fields in an Anti-de-Sitter (AdS) spacetime to a strongly coupled d-dimensional quantum field theory on its conformal boundary \cite {Maldacena:1997re,Gubser:1998bc,Witten:1998qj,Aharony:1999ti}.  Apart from diverse other applications this holographic duality may be utilized to study strongly coupled quantum field theories describing condensed matter systems.  In this context, it was first shown by Gubser \cite{Gubser:2008px} that for a charged AdS black hole minimally coupled to a complex scalar field allows the condensation of the scalar field near the black hole horizon resulting in {\it scalar hair} at a certain critical temperature. From the holographic dictionary this corresponds to  a scalar operator that is dual to the bulk charged scalar field, acquiring a non zero vacuum expectation value in the strongly coupled boundary field theory. The formation of such a charged condensate describes a superconducting phase transition in the strongly coupled boundary quantum field theory that spontaneously breaks the global $U(1)$ symmetry and is referred to as a {\it holographic superconductor} \cite{Hartnoll:2008vx,Hartnoll:2008kx,Herzog:2009xv,Horowitz:2010gk}. Subsequently there was a surge of interest in the investigation of the condensate formation, transport and spectral properties for such holographic superconductors in various dimensions both in the probe limit and including the backreaction \cite{Gregory:2009fj,Ren:2010ha,Brihaye:2010mr,Barclay:2010up,Domenech:2010nf,Chaturvedi:2013ova}. Furthermore in \cite{Franco:2009if, Maeda:2009wv, Herzog:2010vz, Zeng:2010zn, Yin:2013fwa} the authors have studied the thermodynamic properties and the critical phenomena of such holographic superconductors and showed that the critical exponents indicate a mean field behavior corresponding to a second order phase transition. 

In a distinct context over the last decade there has been important progress in associating an intrinsic  Riemannian geometrical structure with equilibrium thermodynamic systems through the studies of  Weinhold \cite{Weinhold:1975sc, Weinhold:1975hp} and Ruppeiner \cite{Ruppeiner:1995zz}. Such a framework of {\it thermodynamic geometry} associates a Riemannian metric with an Euclidean signature in the equilibrium state space of any thermodynamic system which is based on the thermodynamic fluctuations. In a Gaussian approximation the probability distribution of such fluctuations was related to the positive definite  invariant line element defined by this geometry. It was shown that the thermodynamic Riemannian scalar curvature encodes the microscopic interactions of the underlying statistical system. Specifically in \cite{Ruppeiner:1995zz}, it was shown through standard scaling and hyperscaling arguments that the thermodynamic scalar curvature is proportional to the {\it correlation volume} of the system and hence diverges at a critical point of second order phase transition. This geometrical framework was used to characterize phase transitions and critical phenomena for diverse thermodynamic systems {\cite {Ruppeiner:1995zz}. Application of this framework to study the thermodynamics and phase transition for AdS black holes have yielded interesting insights \cite {Sarkar:2008ji, Ruppeiner:2007up, Cai:1998ep, Aman:2003ug, Aman:2006kd, Gibbons:2004ai, Shen:2005nu,Sahay:2010wi,Sahay:2010tx,Sahay:2010yq, Banerjee:2011cz,Kubiznak:2012wp}. Naturally the direct connection between the thermodynamic scalar curvature and the microscopic correlation length makes this framework suitable to study the phase transitions in systems lacking a precise and complete microscopic statistical description like black holes or  strongly coupled condensed matter systems.

In this article we propose to investigate the phase transition and critical phenomena for strongly coupled holographic superconductors in a grand canonical ensemble using the framework of thermodynamic geometry. Quite obviously a direct computation of the thermodynamic geometry for such strongly coupled finite temperature field theories would be intractable. However the gauge/gravity correspondence provides a holographic approach to the problem through the weakly coupled dual bulk gravitational theory. To this end we analytically compute the holographic free energy for the strongly coupled boundary quantum field theory at a finite temperature from the dual bulk charged AdS black hole in presence of charged scalar fields. For this
we utilize an analytic method to implement the bulk to boundary correspondence through a saddle point approximation and in the probe limit as described in \cite{Gregory:2009fj,Barclay:2010up}. We emphasize here that our analytic approach is distinct from the conventional analytic and numerical approach for the computation of the holographic free energy \cite{Herzog:2010vz,Yin:2013fwa}. 

The holographic free energy may then be used as the thermodynamic potential to compute the thermodynamic metric and  the corresponding thermodynamic scalar curvature for the strongly coupled boundary field theory at a finite temperature through standard techniques of Riemannian geometry. The study of the thermodynamic scalar curvature as a function of the temperature then exhibits a divergence at the critical transition temperature for the superconducting phase transition in the boundary field theory for different values of the mass of the bulk charged scalar field. As mentioned earlier such a divergence indicates a critical point of second order thermal phase transition. The critical temperature  for this divergence matches well with the critical temperature obtained from the conventional analytical and numerical techniques based on the condensate formation. There has been no previous such attempt to characterize the phase structure of such a strongly coupled field theory at a finite temperature using the framework of thermodynamic geometry in a holographic approach. We emphasize here that our analytical approach using a geometrical framework based on microscopic fluctuations to study the phase transition and critical phenomena is more elegant and accurate than the conventional approach based on the superconducting condensate formation. This is indicated by the slight difference in the critical temperatures arrived at through the two distinct techniques mentioned here.

This article is organized as follows, in Section 2 we briefly describe the gravitational dual of a holographic superconductor and describe  the superconducting phase of the (2+1) dimensional boundary field theory. Furthermore in the same section we present the computation of the holographic free energy of the strongly coupled (2+1) dimensional boundary field theory. In Section 3 we obtain the thermodynamic metric using the holographic free energy and compute the corresponding thermodynamic scalar curvature for the (2+1) dimensional boundary field theory and study its behavior with the temperature. In Section 4 we present a summary of our results and discuss future open problems.

\section{The Gravity Dual of a Holographic Superconductor}
The minimal model for obtaining a holographic superconductor requires a $U(1)$ gauge field and a charged complex scalar field in an AdS black hole background\cite{Hartnoll:2008vx}.  The bulk action corresponding to the gravitational dual may be given as
\begin{eqnarray}
S &=& \int{d^4 x} \sqrt{-g}\biggl[\frac{1}{2\kappa^2}(R +\frac{6}{L^2})-\frac{1}{4}F^{\mu\nu}F_{\mu\nu}\nonumber\\
  &~&- \frac{1}{2}|\nabla \Psi- i q A\Psi|^2- \frac{1}{2}m^2|\Psi |^2 \biggr]\label{actiom}
\end{eqnarray}
where, $\kappa^2= 4 \pi G_4$ is related to the gravitational constant in the bulk and $L$ is the $AdS$ radius which we set to unity for further analysis. Here, $\Psi$ is the complex scalar field which is charged under the bulk Maxwell field $A_{\mu}$. The constants $q$ and $m$ correspond respectively to the charge and the mass of the bulk scalar field $\Psi$.  Here, we work in a weak gravity (or probe) limit, $q \rightarrow \infty$ in which gravity decouples from the Abelian-Higgs sector (the scalar and the gauge field). In this limit, we consider the  background to be given by a  planar Schwarzschild black hole in the $AdS_4$ bulk with the metric
\begin{eqnarray}
ds^2 &=& \frac{1}{z^2}\left(-  f(z)dt^2 + \frac{dz^2}{f(z)} +dx^2 + dy^2\right),\label{metric}\\
f(z) &=& 1-\frac{z^3}{z_h^3},~~z_h=M^{-1/3}.
\end{eqnarray}
Here, $M$ stands for the mass of the black hole and the points $z=z_h$ , $z\rightarrow 0$ respectively correspond to the horizon and  boundary of asymptotically Anti-de Sitter  space-time. The hawking temperature of the black hole is given as
\begin{equation}
T_h=\frac{\arrowvert f'(z_h)\arrowvert}{4\pi} = \frac{3  }{4\pi z_h } .
\end{equation}

Assuming the ansatz $A_\mu=(\phi(z),0,0,0)$ and $\Psi=\psi(z)$ for the bulk fields\cite{Hartnoll:2008vx}, the equations of motion for the gauge field  and the charged complex scalar field in the background (\ref{metric})  may be expressed as follows \footnote{To be exact, we consider $\Psi=\psi(z)e^{i\alpha}$ and make a gauge transformation $A_{\mu}\rightarrow A_{\mu}+\nabla_{\mu}\alpha$, which renders the equations of motion free from the phase $\alpha$. }
\begin{eqnarray}
\psi''+\left(-\frac{2}{z}+\frac{f'}{f}\right) \psi '+\left(\frac{\phi^2}{f^2}-\frac{m^2}{z^2 f}\right) \psi &=&0,\label{psieom}\\
\phi''-\frac{2 \psi^2 }{z^2 f}\phi &=& 0,\label{phieom}
\end{eqnarray}
where  prime denotes derivative with respect to $z$. An exact solution to equations (\ref{psieom}) and (\ref{phieom}) is clearly $\psi=0$ and $\phi=\mu-\rho~z$ , which corresponds to the normal phase of the  strongly coupled (2+1) dimensional boundary field theory at finite temperature with $\rho$ and $\mu$ as the charge density and the chemical potential respectively.

\subsection{Superconducting Phase}
In this section, we study the superconducting phase of the (2+1) dimensional strongly coupled boundary field theory in the probe limit.  It was observed in \cite{Gubser:2008px}, that a bulk charged AdS black hole develops an instability which leads to the formation of scalar hair near the horizon at low temperatures. This  phase is described by the bulk solution, $\psi\neq 0$ of the equations of motion (\ref{psieom})  and (\ref{phieom}). In the boundary field theory, this corresponds to a superconducting phase transition with a charged scalar operator ${\cal O}$ dual to $\psi$ acquiring a non zero vacuum expectation value at the critical temperature. 

From the equations of motion (\ref{psieom})  and (\ref{phieom}), we observe that for a nontrivial solution we need to determine the two independent functions $(\psi(z),\phi(z))$. For this suitable boundary conditions must be imposed at the conformal boundary $z\rightarrow 0$ and at the black hole horizon $z=z_h$ in the $AdS_4$ bulk.  For a regular event horizon $z=z_h$, we have the boundary conditions as 
\begin{eqnarray}
f(z_h)=0,~\phi(z_h)=0,~\psi'(z_h)=-\frac{ m^2}{3}\psi(z_h),\label{bchorizon}
\end{eqnarray}
so that the term $g^{\mu\nu} A_{\mu}A_{\nu}$ may remain finite at the horizon \cite{Gubser:2008px}.  We also require the functions $\{\psi(z),\phi(z)\}$ to admit finite values and Taylor series expansions near the horizon as,
\begin{eqnarray}
\psi_h(z) &=&\psi(z_h)+\psi'(z_h)(z-z_h)\nonumber\\
&&+\frac{\psi''(z_h)}{2}(z-z_h)^2+\cdots, \label{psih}\\
\phi_h(z) &=&-\phi'(z_h)(z-z_h)\nonumber\\
&&+\frac{\phi''(z_h)}{2}(z-z_h)^2+\cdots. \label{phih}
\end{eqnarray}

Next following \cite{Gregory:2010yr}, we compute the undetermined coefficients $\phi''(z_h)$ and $\psi''(z_h)$ using equations of motion (\ref{psieom}, \ref{phieom}) and horizon expansions (\ref{psih}, \ref{phih}) as
\begin{eqnarray}
\phi''(z_h)&=&\frac{2}{3}\phi'(z_h)\psi(z_h)^2,\label{phi2}\\
\psi''(z_h)&=&-\frac{\psi(z_h)}{18z_h^2}(m^4+6m^2-z_h^4\phi'(z_h)^2).\label{psi2}
\end{eqnarray}

Using (\ref{phi2}) and (\ref{psi2}) we may write down the modified near horizon expansion of $\psi$ and $\phi$ upto second order in the derivatives as
\begin{eqnarray}
\psi_h(z) &=&\frac{\psi(z_h)}{36 z_h^2}(z_h^4\phi'(z_h)^2-m^4-6m^2)(z-z_h)^2\nonumber\\
&&-\frac{m^2}{3z_h}\psi(z_h)(z-z_h)+\psi(z_h)\label{psihorz}\\
\phi_h(z)&=&\phi'(z_h)(z-z_h)\Bigl(\frac{\psi(z_h)^2}{3}(z-z_h)-1\Bigr).\label{phihorz}
\end{eqnarray}

Thus we are left with only three positive independent parameters at the horizon $\{z_h,\psi(z_h),\phi'(z_h)\}$.  Furthermore, the asymptotic form of the functions $\{\psi(z),\phi(z)\}$ near the $AdS$ boundary $z\rightarrow 0$ may be written as,
\begin{eqnarray}
\psi_b(z) &=& \psi_{-}~z^{\Delta_{-}} +\psi_{+}~z^{\Delta_{+}} +\cdots,\nonumber\\
\phi_b(z) &=&\mu-\rho~z+\cdots,\label{bdryexp}
\end{eqnarray}
where, $\Delta_{\pm}=\frac{3}{2}\pm\sqrt{\frac{9}{4}+m^2}$.  Here  the coefficients $\mu,\rho,\psi_{-}=<{\cal O}_{-}>$ and $\psi_{+}=<{\cal O}_{+}>$ represent the chemical potential, charge density, source and vacuum expectation value of the dual charged scalar operator ${\cal O}$ respectively in the dual boundary field theory. It is to be noted that for, $-5/4>m^2>-9/4$ the mass of the scalar field lies near the B-F (Breitenlohner-Freedman) bound \cite{Breitenlohner:1982bm} which renders both the modes $\psi_{-}$ and $\psi_{+}$ normalizable whereas, for $-5/4<m^2$ only the mode $\psi_{+}$ is normalizable .Thus one may impose the condition that either $\psi_{-}$ or $\psi_{+}$ vanish at the $AdS$ boundary $z\rightarrow 0$. We assume $(\psi_{-}=0,\psi_{+}\neq 0 )$ which reflects that the condensate $\psi_{+}=<{\cal O}_{+}>$ arises spontaneously in the boundary field theory in the absence of  sources.

\begin{figure}[H]
\centering
\includegraphics[width =2.1in,height=1.3in]{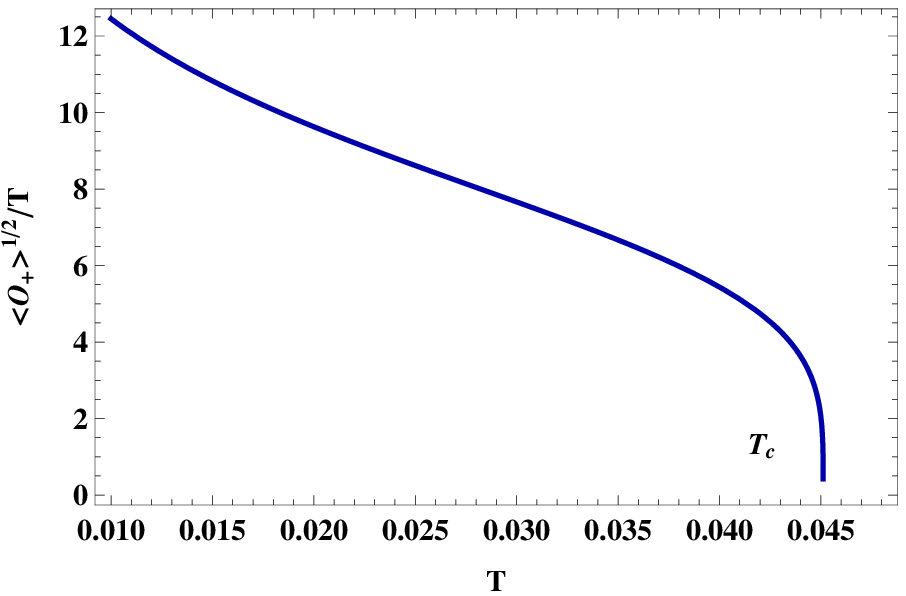}
\\
\includegraphics[width =2.1in,height=1.3in]{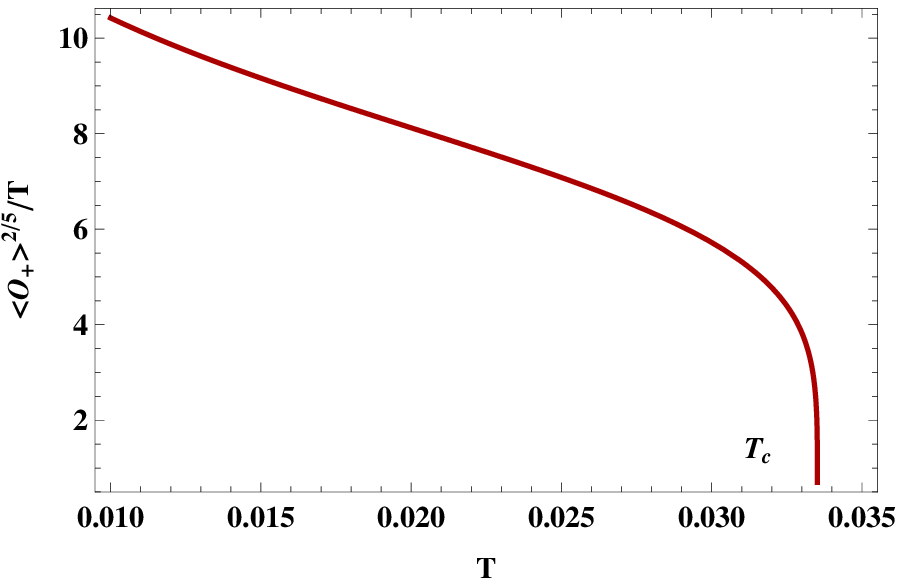}
\\
\includegraphics[width =2.1in,height=1.3in]{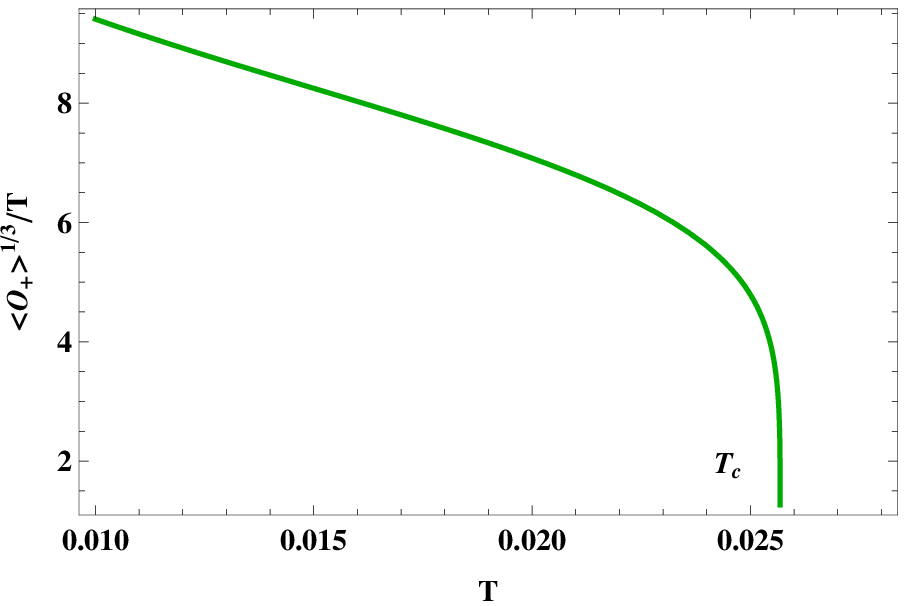}
\caption{\label{fig:conds1} The condensate $<{\cal O}_{+}>^{1/\Delta}/T$ is plotted with respect to the temperature $T$ for fixed value of $\mu=1$. The graphs from top to bottom correspond to values $m^2=-2,-5/4,0$ of the mass of the scalar field with the critical temperatures  $T_c=0.045,0.034,0.026$ respectively. } \end{figure} 
To obtain the condensate $\psi_{+}=<{\cal O}_{+}>$, we begin with sewing the horizon and the boundary expansions (\ref{psihorz}, \ref{phihorz}, \ref{bdryexp}) of the fields $\phi$ and $\psi$ near $z=z_h/2$. We also match the derivatives of the boundary and the horizon expansions for the fields near the sewing point as
\begin{eqnarray}
\phi_h\Bigl(\frac{z_h}{2}\Bigr) &=&\phi_b\Bigl(\frac{z_h}{2}\Bigr),~\phi'_h\Bigl(\frac{z_h}{2}\Bigr)=\phi'_b\Bigl(\frac{z_h}{2}\Bigr),\label{phim}\\
\psi_h\Bigl(\frac{z_h}{2}\Bigr) &=&\psi_b\Bigl(\frac{z_h}{2}\Bigr),~\psi'_h\Bigl(\frac{z_h}{2}\Bigr)=\psi'_b\Bigl(\frac{z_h}{2}\Bigr).\label{psim}
\end{eqnarray}

From Eqs. (\ref{phim})  we arrive at following relations,
\begin{eqnarray}
\psi(z_h)=2 \sqrt{\frac{\mu}{z_h\phi'(z_h)} -1},~\rho=\frac{4 \mu -\phi'(z_h) z_h}{3 z_h}.\label{phimrel}
\end{eqnarray}
Similarly, from Eqs. (\ref{psim})  we obtain
\begin{eqnarray}
\phi'(z_h)&=&\frac{\sqrt{144 \Delta +(\Delta +2) m^4+6 (5 \Delta +6) m^2}}{z_h^2\sqrt{\Delta +2} },\nonumber\\
\psi_{+}&=&\frac{2^{\Delta -1} \left(m^2+12\right) \psi(z_h)}{3 (\Delta +2) z_h^{\Delta }} ,\label{psimrel}
\end{eqnarray}
where, $\Delta=\Delta_{+}$. Now using Eqs. (\ref{phimrel}) and (\ref{psimrel}) with the expression for the Hawking temperature $T=3/4\pi z_h$, the expression of $<{\cal O}_{+}>$  may be expressed as a function of temperature $T$ and charge density $\rho$. As an example, the expression of $<{\cal O}_{+}>$ with $m^2=-2$ (mass of scalar field near BF bound) is as follows
\begin{eqnarray}
<{\cal O}_{+}> &=&\sqrt{2}\psi_{+}=\frac{160\pi ^2 }{27}\sqrt{2T^3 T_c} \sqrt{1-\frac{T}{T_c}},\label{conds}\\
T_c &=& \frac{3\mu}{8\pi \sqrt{\sqrt{7}}}=0.0451\mu.\label{tcritical}
\end{eqnarray}
In fig.(\ref{fig:conds1}) we have further plotted the dimensionless quantity $<{\cal O}_{+}>^{1/\Delta}/T$ with respect to temperature $T$ for different values of $m^2=(-2, -5/4, 0)$ at fixed $\mu=1$ in the grand canonical ensemble. From the eq.(\ref{conds}) and fig.(\ref{fig:conds1}), we observe that $<{\cal O}_{+}>$ goes to zero at the critical temperature $T=T_c$ and the condensate only exists for $T<T_c$. In the limit $T\rightarrow T_c$, we also recover the mean field result $<{\cal O}_{+}>\approx (T-T_c)^{1/2}$ indicating a second order phase transition in the Landau-Ginzburg framework as observed in \cite{Herzog:2010vz}. 

\subsection{Holographic Free Energy}
In this section we obtain the free energy of the strongly coupled (2+1) dimensional boundary field theory at a finite temperature
using a holographic approach \cite{Hartnoll:2008vx}. Through the AdS/CFT dictionary the free energy ($\Omega$) of the boundary field theory is related to the product of temperature ($T$) and the on-shell Euclidean action $S_E$. Furthermore proper boundary terms must be added to the Euclidean action in order to define a well posed variational problem. These involve the usual boundary cosmological constant and the Gibbons Hawking terms required for regulating the Einstein-Hilbert action in an asymptotically AdS space-time \cite{PhysRevD.15.2752}.  The Euclidean action $S_E$ with proper boundary terms may be described as follows
\begin{eqnarray}
-S_{E} &=&\int{d^4x}\sqrt{-g}\left[\frac{1}{2\kappa^2}(R+6)+{\cal L}_m\right]\nonumber\\
&~&+\frac{1}{2\kappa^2} \int_{z\rightarrow 0}{d^3x}\sqrt{-h}(2{\cal K}-4).\label{Eaction}
\end{eqnarray}
Here, ${\cal L}_m$ refers to the matter Lagrangian (Abelian-Higgs sector), $h$ is the determinant of the induced metric on the $AdS$ boundary and ${\cal K}$ is the trace of the extrinsic curvature ${\cal K}_{\mu\nu}$. 

As described earlier, we are working in the probe limit which makes it feasible to  relate the free energy of the boundary field theory to the on-shell value ($S_{os}$) of the Abelian-Higgs sector of the Euclidean action $S_E$ \cite{Herzog:2010vz}. Now using the equations of motion (\ref{psieom}, \ref{phieom}) and the metric (\ref{metric}),  the on shell action $S_{os}$ in a saddle point approximation may be written down as
\begin{eqnarray}
S_{os} &=&\int{d^3x}\biggl[ \frac{f(z)\psi(z)\psi'(z)}{2z^2}\Big|_{z=0} \nonumber\\
&-&\frac{1}{2}\phi(z)\phi'(z)\Big|_{z=0}-\frac{1}{2}\int_0^1{dz}\frac{\phi(z)^2\psi(z)^2}{z^2f(z)}\biggr].\label{onshell1}
\end{eqnarray}

Using the asymptotic forms of $\phi$ and $\psi$ from eq.(\ref{bdryexp}) in (\ref{onshell1}), we obtain
\begin{eqnarray}
S_{os} &=&\int{d^3x}\biggl[ \frac{\mu\rho}{2}+\frac{3\psi_{+}\psi_{-}}{2}+\left(\frac{\psi_{-}^2}{2z}\right)\bigg|_{z=0} \nonumber\\
&-&\frac{1}{2}\int_0^1{dz}\frac{\phi(z)^2\psi(z)^2}{z^2(1-z^3/z_h^3)}\biggr].\label{onshell2}
\end{eqnarray}

In addition, for carrying out a holographic renormalization of the free energy it is also required to add a boundary counter term $(S_{ct})$ to the action as described in \cite{Herzog:2010vz}. Using the expressions of $\phi$ and $\psi$ from eq.(\ref{bdryexp}), the boundary counter term $S_{ct}$ may be given as follows\footnote{It is to be noted that here, we have chosen the condition $\{\psi_{-}=0,\psi_{+}\neq 0\}$ at the $AdS$ boundary $z=0$. However, in order to implement the condition $\{\psi_{-}\neq 0,\psi_{+}= 0\}$ one must consider a boundary counter term $S_{ct}=\int{d^3x}\sqrt{-h}(\frac{1}{z^4} \psi(z)\psi'(z))|_{z= 0}$ as described in \cite{Herzog:2010vz}. }
\begin{eqnarray}
S_{ct} &=&-\frac{1}{2}\int{d^3x}(\sqrt{-h} |\psi(z)|^2)\Big|_{z= 0}\nonumber\\
&=&-\frac{1}{2}\int{d^3x}\biggl[\frac{\psi_{+}\psi_{-}}{2}+\left(\frac{\psi_{-}^2}{2z}\right)\bigg|_{z=0}\biggr].\label{counter}
\end{eqnarray}

Now using equations (\ref{onshell2}) and (\ref{counter}), we find the expression for the free energy of the boundary field theory in the grand canonical ensemble \footnote{Note that, one can also work with the free energy in canonical ensemble as described in\cite{Herzog:2010vz}}. However, in this paper we restrict ourselves only to the expression of free energy in the grand canonical ensemble as
\begin{eqnarray}
\Omega=- T(S_{os}+S_{ct})= \beta T V_2\biggl[-\frac{\mu\rho}{2}-\frac{\psi_{+}\psi_{-}}{2} +{\cal I}\biggr].\label{FreeEn}
\end{eqnarray}
where, $\int{d^3x}=\beta V_2$ and the integral ${\cal I}$ is given by the expression
\begin{equation}
{\cal I}=\frac{1}{2}\left(\int_0^{1/2}{dz}+\int_{1/2}^1{dz}\right)\frac{\phi(z)^2\psi(z)^2}{z^2(1-z^3/z_h^3)}
\end{equation}

In order to determine an analytic expression for the free energy it is required to compute the integral ${\cal I}$. For this, we replace $\phi$ and $\psi$ by their horizon expansions (\ref{phihorz}) and (\ref{psihorz}) respectively between the limits $1$ and $1/2$. Similarly  between the limits $0$ and $1/2$, we replace $\phi$ and $\psi$ by their boundary expansions (\ref{bdryexp}). This leads to an analytic expression of ${\cal I}$ in term of variables $z_h,\mu,\rho,m,\psi_{-},\psi_{+},\phi'(z_h)$ and $\psi(z_h)$. From equations (\ref{phimrel}), (\ref{psimrel}) and (\ref{FreeEn})  it is clear that the analytic expression of free energy with the $AdS$ boundary condition $\psi_{-}=0$ and mass $m^2=-2$ of the scalar field, may be expressed as a function of $z_h$ and $\mu$  as 
\begin{eqnarray}
\frac{\Omega}{V_2}&=&0.029\mu ^3 -\frac{0.713\mu ^2}{z_h}+\frac{0.549\mu }{z_h^2}-\frac{1.259}{z_h^3}.\label{freemurh}
\end{eqnarray}
\begin{figure}[H]
\centering
\includegraphics[width =2.5in,height=1.5in]{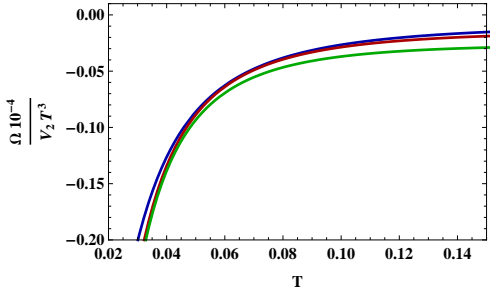}
\caption{\label{fig:FreeE}  The quantity $\frac{\Omega\times 10^{-4}}{T^3V_2}$  is plotted with respect to the temperature $T$ for fixed value of $\mu=1$. The blue, red and green curves correspond to values $m^2=-2,-5/4,0$ of the mass of the scalar field . } \end{figure} 
The free energy for different values of the mass $m$ of the charged scalar may be obtained in a similar fashion as described above. Further substituting,  $z_h=\frac{3}{4\pi T}$ in (\ref{freemurh}) we obtain the free energy in the grand canonical ensemble as a function of temperature $T$ and chemical potential $\mu$ as 
\begin{eqnarray}
\frac{\Omega}{V_2}=0.029 \mu ^3-92.567 T^3+9.648 \mu  T^2-2.987 \mu ^2 T.\label{FreeTmu}
\end{eqnarray}

Using equations (\ref{phimrel}),(\ref{psimrel}) and (\ref{freemurh}), we also compute the free energy of the  (2+1) dimensional boundary field theory as a function of temperature $T$ and charge density $\rho$ as,
\begin{eqnarray}
\frac{\Omega}{V_2}&=&1.673\times 10^{-4}\frac{\rho ^3}{T^3}-\frac{0.080 \rho ^2}{T}\nonumber\\
&&-125.872 T^3-3.719 \rho  T.\label{FreeTrho}
\end{eqnarray}

In fig.(\ref{fig:FreeE}), we have plotted the free energy density  $\omega=\Omega/V_2$ against the temperature for various values of the mass of the scalar field at fixed value of chemical potential $\mu=1$ . It is observed from the plots that the free energy decreases as the temperature is increased indicating formation of condensate below a certain critical temperature $T_c$ indicating a second order superconducting phase transition in the boundary field theory. Furthermore, we have also  computed the free energy density  $\omega=\Omega/V_2$  numerically which compares well with the analytical expression for the free energy \cite{Hartnoll:2008vx}.

\section{Thermodynamic geometry}
In this section we obtain the thermodynamic geometry and the consequent scalar curvature in the grand canonical ensemble
for the holographic superconductors described in the previous section . For the strongly coupled boundary field theory in a grand canonical ensemble it is suitable to use the free energy representation due to Ruppeiner \cite {Ruppeiner:1995zz} for computing the thermodynamic metric. For a particular fixed value of the mass of the scalar field,  $m^2=-2$  the thermodynamic metric may be obtained  as a Hessian of the Gibb's free energy density $\omega=\Omega/V_2$  (\ref{FreeTrho}) of the (2+1) dimensional boundary field theory with respect to the temperature $T$ and the charge density $\rho$ as
\begin{equation}
g_{ij}=-\frac{1}{T} \frac{\partial^2 \omega(T,\rho)}{\partial x^i \partial x^j},\label{metricG}
\end{equation}

where, $x_i=(T,\rho )$. From (\ref{metricG}) the independent components of the thermodynamic metric may be written down as follows
\begin{eqnarray}
g_{TT} &=& -\frac{0.002 \rho ^3}{T^6}+\frac{0.161 \rho ^2}{T^4}+755.232,\nonumber\\
g_{T\rho}=g_{\rho T}&=&\frac{0.002 \rho ^2+3.719 T^4-0.161 \rho  T^2}{T^5},\nonumber\\
g_{\mu\mu}&=&\frac{0.161 T^2-0.001 \rho }{T^4}.\label{MetricG}
\end{eqnarray}

From the expression for the thermodynamic metric in (\ref{MetricG}) it is straightforward to obtain the thermodynamic scalar curvature $R$ through standard techniques of Riemannian geometry \cite{Ruppeiner:1995zz} as follows
\begin{eqnarray}
R&=&-\frac{1}{\sqrt{g}}\Bigl[\frac{\partial}{\partial T}\Bigl(\frac{g_{T\rho}}{g_{TT}\sqrt{g}}\frac{g_{TT}}{\partial \rho}-\frac{1}{\sqrt{g}}\frac{g_{\rho\rho}}{\partial T}\Bigr)\nonumber\\
&+&\frac{\partial}{\partial \rho}\Bigl(\frac{2}{\sqrt{g}}\frac{g_{T\rho}}{\partial \rho}-\frac{1}{\sqrt{g}}\frac{g_{TT}}{\partial \rho}-\frac{g_{T\rho}}{g_{TT}\sqrt{g}}\frac{g_{TT}}{\partial T}\Bigr)\Bigr].\label{Rcurvg}
\end{eqnarray}

Here, $g$ stands for the determinant of the thermodynamic metric. From (\ref{Rcurvg}), the expression for thermodynamic scalar curvature $R$ may be obtained as a function of the temperature $T$ and the charge density $\rho$. However, for the grand canonical ensemble considered by us it is required to investigate the thermodynamic scalar curvature $R$ as a function of the temperature $T$ a fixed value of the chemical potential $\mu$. Thus using the equations  (\ref{phimrel}) and (\ref{psimrel}) with $z_h=\frac{3}{4\pi T}$ we obtain the expression for $R$ as a function of temperature $T$ and the chemical potential $\mu$  as follows \footnote{ The thermodynamic scalar curvature $R$ as a function of  $T$ and the chemical potential $\mu$ for different values of the mass $m$ of the scalar field may be obtained in a similar way as described above. }
\begin{eqnarray}
R&=&-\frac{{\cal N}}{{\cal D}^3},\label{RuppC}\\
{\cal N}&=&-0.002 T^{10}+7.856\times 10^{-5} \mu  T^9\nonumber\\
&&+2.808 \times 10^{-5}\mu ^2 T^8-7.218\times 10^{-7} \mu ^3 T^7\nonumber\\
&&-2.711\times 10^{-8} \mu ^4 T^6+7.373 \times 10^{-10}\mu ^5 T^5\nonumber\\
&&-3.058\times 10^{-11} \mu ^6 T^4+5.616\times 10^{-13} \mu ^7 T^3\nonumber\\
&&-6.816\times 10^{-27} \mu ^8 T^2+8.863\times 10^{-29}\mu ^9 T,\label{RuppCN}\\
{\cal D}&=&T^4+0.078 \mu  T^3-0.005 \mu ^2 T^2\nonumber\\
&&+6.553 \times 10^{-5} \mu ^3 T-2.956\times 10^{-6} \mu ^4 .\label{RuppCD}
\end{eqnarray}
The temperature at which the thermodynamic scalar curvature diverges is determined from the zero of the denominator (\ref{RuppCD}) of the expression (\ref{RuppC}) as $T_c=0.0424 \mu$. This critical temperature matches well with that obtained through a conventional approach based on the condensate formation as given in  eq.(\ref{tcritical}). Thus we observe that for the value $m^2=-2$ of the mass of the scalar field, the thermodynamic scalar curvature diverges at the temperature $T_c= 0.0424 \mu$ indicating a second order superconducting phase transition \cite{Ruppeiner:1995zz} in the strongly coupled boundary field theory . In fig.(\ref{fig:RvsT}) below we have plotted the thermodynamic scalar curvature $R$ against the temperature for different values of the mass of the scalar field $m^2= (-2, -5/4, 0)$ for a fixed value of the chemical potential $\mu=1$. The graphs clearly illustrate the divergence of the thermodynamic scalar curvature at a critical transition temperature for the different masses of the bulk charged scalar field which characterizes the superconducting phase transition.

\begin{figure}[H]
\centering
\includegraphics[width =2.3in,height=1.5in]{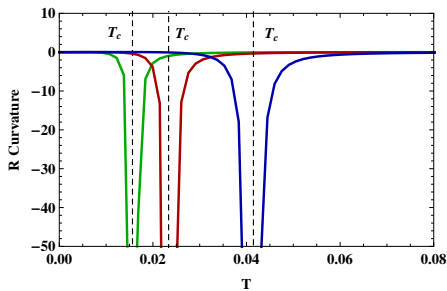}
\caption{\label{fig:RvsT} The thermodynamic scalar curvature $R$ is plotted with respect to the temperature $T$ for fixed value of $\mu=1$. The blue, red and green curves correspond to values $m^2=-2,-5/4,0$ of the mass of the scalar field with the scalar curvature diverging at $T_c=0.0424,0.024,0.016$ respectively. } \end{figure} 

 For a comparison we have also obtained the thermodynamic scalar curvature from a numerical computation of the holographic free energy \cite{Hartnoll:2008vx} which exhibits a similar divergent behavior with temperature as described in fig.(\ref{fig:RvsT}). 
As mentioned in the introduction our analytical method using geometrical framework provides a more elegant and accurate characterization of the superconducting phase transition for the strongly coupled boundary field theory than the conventional approach based on the condensate formation. This explains the marginal difference in the values of the critical temperature 
$T_c$  obtained from the divergence of the thermodynamic scalar curvature $R$ and that determined from the zero of the condensate $<{\cal O}_{+}>$ in (\ref{conds}).

\section{Summary and Discussions}

In summary we have obtained  the thermodynamic geometry of a strongly coupled (2+1) dimensional quantum field theory at a finite temperature in a holographic set up using the gauge/gravity correspondence.  The corresponding thermodynamic scalar curvature was computed and used to characterize the superconducting phase transition in the strongly coupled boundary field theory at a finite temperature. We emphasize here that our study is the first such attempt to characterize the superconducting phase transition for a strongly coupled boundary field theory in a holographic scenario through the framework of thermodynamic geometry.  The (2+1) dimensional strongly coupled boundary field theory is holographically dual to a charged bulk (3+1) dimensional AdS black hole in presence of charged scalar fields. In this context we have analytically computed  the holographic free energy of the (2+1) dimensional boundary field theory in a grand canonical ensemble from the gravitational dual through the bulk boundary correspondence in the probe limit. The thermodynamic metric was then computed from the Hessian of the holographic free energy with respect to the temperature $T$ and the charge density $\rho$  in a standard fashion \cite{Ruppeiner:1995zz}. The associated thermodynamic scalar curvature for the strongly coupled boundary field theory could then be obtained from this metric using standard techniques of  Riemannian geometry. 

The superconducting phase transition and critical phenomena for the strongly coupled boundary field theory in a grand canonical ensemble was then investigated through the variation of the thermodynamic scalar curvature as a function of the temperature with a fixed chemical potential for different masses of the bulk charged scalar field. Remarkably the thermodynamic scalar curvature obtained holographically diverges at the critical transition temperature indicating a critical point of second order  phase transition in the strongly coupled boundary field theory. The critical temperature for this superconducting phase transition at which the scalar curvature diverges compares well with the critical temperature obtained from earlier numerical as well as analytic studies of the condensate formation. Our approach employing an analytical and geometrical framework to study the critical phenomena for holographic superconductors is clearly more elegant and accurate than the conventional approach in \cite {Hartnoll:2008vx,Herzog:2010vz,Yin:2013fwa}.  As mentioned earlier this accounts for the marginal difference in the values for the critical temperatures obtained through the two distinct techniques. Our investigations clearly and directly illustrate that the superconducting phase transition describing a holographic superconductor is a critical point of a second order phase transition which is also consistent with the mean field nature of the result.

An important open problem for future investigation is to study the phase structure of other more complex holographic superconductors in the geometrical framework described by us. In this context it would be interesting to study the phase structure of the {\it generalized holographic superconductors} \cite{Franco:2009yz} involving a first order superconducting phase transition in the boundary field theory.  It would also be interesting to study the phase structure of the {\it p-wave} and {\it d-wave} holographic superconductors \cite{Gubser:2008wv,Cai:2013aca,Chen:2010mk} using our techniques. Another future issue would be to investigate the thermodynamic geometry of holographic superconductors in an external magnetic field which is dual to a charged dyonic AdS black hole in presence of charged scalar fields \cite{Albash:2008eh,PhysRevLett.103.091601,Ge:2010aa}. We hope to return to these interesting issues in the future.

\section{Acknowledgment}
This work of Pankaj Chaturvedi is supported by grant no. 09/092(0846)/2012-EMR-I from CSIR India.

\bibliography{TG_holo_sup}
\bibliographystyle{unsrt}

\end{document}